\documentclass[twocolumn,aps,prl,reprint,superscriptaddress,floatfix]{revtex4-2}
\usepackage{amsmath}
\usepackage{graphicx}
\usepackage{tabularx}
\usepackage{comment}
\usepackage{booktabs} 
\usepackage{lineno}
\usepackage{soul}
\usepackage[table,xcdraw]{xcolor}
\usepackage{colortbl} 
\usepackage{placeins}
\usepackage{makecell}
\usepackage{hyperref}
\usepackage{makecell,multirow}
\usepackage{siunitx}
\usepackage{array}

\definecolor{headerblue}{rgb}{0.0, 0.2, 0.5} 
\definecolor{rowblue1}{rgb}{0.8, 0.85, 1.0}  
\definecolor{rowblue2}{rgb}{0.9, 0.95, 1.0}  
\begin{document}

\renewcommand{\thesection}{\arabic{section}} 
\renewcommand{\thefigure}{\arabic{figure}} 
\renewcommand{\thetable}{\arabic{table}}

\newcommand{\invcm}{\unit{cm^{-1}}}
\newcommand{\icm}{\unit{cm^{-1}}}
\newcommand{\etg}{E$_{2g}$}
\newcommand{\aog}{A$_{1g}$}
\newcommand{\eog}{E$_{1g}$}
\newcommand{\BT}[1]{\textcolor{teal}{BT: #1}}
\newcommand{\IM}[1]{\textcolor{blue}{IM: #1}}
\newcommand{\KB}[1]{\textcolor{olive}{KB: #1}}

\begin{abstract}
MnTe has recently attracted exceptional attention due to its well-established altermagnetism, prompting a thorough reexamination of its properties. In particular, it was found that a Raman-active excitation at \qty{~175}{\per\centi\meter}, routinely assigned to the $E_{2g}$ phonon, is incompatible with this interpretation. 
It was further hypothesized that this
mode is a ``leakage'', due to symmetry lowering, of an otherwise forbidden phonon. Here, using first-principles calculations, we decisively rule out this hypothesis and propose an alternative interpretation that the ``mystery mode'' is an electronic excitation, i.e., a plasmon, enabled by hole self-doping. The resolution of this mystery will require additional experiments and shed new light on the nature of electronic transport in MnTe.
\end{abstract}

\title{Mystery of the $\mathbf{\qty{175}{\per\centi\meter}}$
Raman Mode in MnTe {Altermagnet}}

\author{Bishal Thapa}
\affiliation{Department of Physics and Astronomy, George Mason University,  Fairfax, VA 22030, USA}
\affiliation{Quantum Science and Engineering Center, George Mason University, Fairfax, VA 22030, USA}

\author{K. D. Belashchenko}
\affiliation{Department of Physics and Astronomy and Nebraska Center for Materials and Nanoscience, University of Nebraska-Lincoln, Lincoln, Nebraska 68588, USA}

\author{Igor I. Mazin}
\affiliation{Department of Physics and Astronomy, George Mason University,  Fairfax, VA 22030, USA}
\affiliation{Quantum Science and Engineering Center, George Mason University, Fairfax, VA 22030, USA}

\date{\today }
\maketitle

\paragraph{Introduction --}
MnTe has gained a lot of attention recently.
Arguably, it is the best studied altermagnetic material, with 25 papers published and 47 preprints posted just in 2025 (see Ref. \cite{rising} and references therein). {One of the reasons for the broad scientific appeal of MnTe is its perceived simplicity. However, as discussed in this Letter, there is a completely unresolved and not understood issue related to its Raman spectrum, which may have important ramifications for its electronic structure and transport properties.}

Despite intense interest in MnTe, only a handful of Raman studies have been published  \cite{Mobasser,Suszkevicz,Szuszpss,Zhang2020,kluczyk2023,wu2025}, as summarized in Table \ref{table}. 
While all papers report a Raman mode around 170--200 cm$^{-1}$, the results are somewhat contradictory in terms of the exact frequency, temperature variation, and polarization dependence. In this respect, the recent preprint of Wu \emph{et al.} \cite{wu2025} stands out. While  previously this mode had been assigned to the $E_{2g}$ phonon (the only one allowed by symmetry), they pointed out that density-functional-theory (DFT) calculations place this phonon below 100 cm$^{-1}$. Such an error is highly unusual in DFT, especially in such a simple material. In Ref. \cite{wu2025} the calculated frequency was 83 cm$^{-1}$; we have verified that varying input parameters, such as the Hubbard $U$, or changing the density functional to metaGGA or hybrid ones, cannot raise this frequency to more that 100 cm$^{-1}$. Furthermore, Wu \emph{et al.} have also measured the polarization dependence and found the mode to be active only in the parallel polarization, while by symmetry {\cite{Bilbao}} 
the $E_{2g}$ phonon should be equally active in both geometries. (This claim contradicts some earlier claims \cite{Suszkevicz,Szuszpss} based on  thin films, which can probably be considered superseded by these more recent measurements.)

\begin{table}[ht]
\centering
\vspace{0.3em}
\begin{tabular}{@{}lcccccl@{}}
\toprule
$\omega$, cm$^{-1}$ & $T$, K & polarization & ref & note & year \\
\midrule
$\sim 178$ & 320 & -- & \multirow{2}{*}{\cite{Mobasser}} & \multirow{2}{*}{bulk} & \multirow{2}{*}{1985} \\
$\sim 187$ & 100 & -- &  &  &  \\
$\sim 215$ & 27 & XY only & \cite{Suszkevicz} & film & 1997 \\
$195-197$ & 25 & XX or XY & \cite{Szuszpss} & film & 2014 \\
$\sim 175$ & 360 & -- & \multirow{2}{*}{\cite{Zhang2020}} & \multirow{2}{*}{film} & \multirow{2}{*}{2020} \\
$\sim 177$ & 200 & -- &  &  &  \\
$\sim 167$ & 310--360 & -- & \multirow{2}{*}{\cite{kluczyk2023}} & \multirow{2}{*}{Xtal} & \multirow{2}{*}{2024} \\
$\sim 170$ & 240 & -- &  &  &  \\
$\sim 175$ & RT & XX only & \cite{wu2025} & Xtal & 2025 \\
$\sim 175$ & 320 & \multirow{2}{*}{*} & \multirow{2}{*}{\cite{coherentph}} & \multirow{2}{*}{Xtal} & \multirow{2}{*}{2021} \\
$\sim 178$ & 77 &  &  &  &  \\
\bottomrule
\end{tabular}
\caption{Available data on the Raman 170--200 mode.\\$^*$ The last entry shows the frequency of a coherent phonon extracted from pump-probe experiments.}
\vspace{0.5em}
\label{table}
\end{table}

Overall, Ref. \cite{wu2025} makes a convincing case against the assignment of the 175 cm$^{-1}$ mode to the $E_{2g}$ phonon, raising the question: What is it then? {A leakage of modes forbidden by symmetry or by momentum conservation is highly unlikely, since it would require a large concentration of symmetry-breaking defects, and is not expected to produce a single narrow line because all momenta contribute to such leakage. No plausible candidate for an impurity phase exists, as opposed to the 120--140 cm$^{-1}$ modes that can be associated with Te precipitates \cite{Szuszpss}.} Ref. \cite{wu2025} proposes an interesting, albeit probably incorrect, as we argue later in this paper, scenario: they suggest that the experimentally well established crystal structure of the space group P6$_3$/mmc (\#194) is incorrect, and in reality the Mn planes are alternatively shifting up and down along $c$ by $\delta\sim \pm 0.001 c\approx 0.007\AA$ - which is 20 times smaller than the zero-point motion amplitude for this mode! While Wu \emph{et al.} correctly point out that in the resulting space group, P\={6}m2 (\#187) the originally silent $B_{1u}$ mode becomes Raman-active (but not in the cross polarization), they missed the point that the acquired activity is proportional to $\delta^2$, so likely negligibly small.

In the following, we carefully verify this hypothesis by (i) fully optimizing the DFT crystal structure, and showing that it invariably converges to the  P6$_3$mmc structure, and (ii) by calculating, using the Placzek formula \cite{Placzek} the Raman activity, and showing that it is negligibly small. We further suggest another explanation, that the mode in question is in fact a plasmon excitation of the self-doped holes. We calculate the plasmon frequencies in DFT and show that they are in the right frequency range. Note that plasmon is only Raman active in the XX polarization, in agreement with the experiment in Ref. \cite{wu2025}.

It remains to be seen why the plasmon frequency (if it is one) is relatively robust from sample to sample. In Table \ref{table3} we summarized reported measurements of the hole concentration $n$. There are some outliers, which also contradict each other, but in general for the stoichiometric MnTe $n\sim 6-11\times 10^{18}$, indicating a variation in $\omega_p$ by $\pm 13$\%, while the frequency of the mode in question at room temperature (RT) varies between different experiments by less than 4\%.  On possible explanation is that  $\omega_p$ is a direct probe of carrier density, regardless of the transport relaxation time, while the Hall coefficient can be dramatically affected by different bands crossing the Fermi level having different relaxation rates, plus can be contaminated by anomalous Hall conductivity.

If our proposal is confirmed by experiment, it will open up another window into the so far poorly understood origin of this hole doping, in particular, why the doping level appears to be rather stable between different samples.

\begin{table}[ht]
\centering
\vspace{0.3em}
\begin{tabular}{@{}lcccccl@{}}
\toprule
$n$, $10^{18}$ cm$^{-3}$ & $T$, K & method & ref & note & year \\
\midrule
50 & 77--400 & Hall & \cite{WASSCHER1964} & bulk & 1964 \\
0.04 & $\sim 0$ & Hall & \multirow{2}{*}{\cite{Xie2024}} & \multirow{2}{*}{bulk} & \multirow{2}{*}{2014} \\
2 & 310 &  &  &  &  \\
6.8 & RT & Hall & \cite{REN2016} & bulk & 2016 \\
3.2 & RT & Hall & \cite{Zulkifal} & bulk Mn$_{1.06}$Te & 2016 \\
$<10$ & 300 & Hall & \cite{zu2017} & bulk & 2017 \\
6 & $\sim 0$ & Hall & \cite{Kriegner2016} & film & 2016 \\
\bottomrule
\end{tabular}
\caption{Available data on the carrier concentration.}
\vspace{0.5em}
\label{table3}
\end{table}

\paragraph{Forbidden-phonon hypothesis --}
As discussed above,  Wu et al.\cite{wu2025} proposed that the true crystal structure belongs to the lower symmetry group $P\bar{6}m2$ (\#187) rather than the commonly reported $P6_3/mmc$ (\#194). 
In this scenario, small alternating displacements of the Mn planes along c axis (by $\delta = \pm 0.1\% c$)  are conjectured to render the otherwise silent B$_{1u}$ mode Raman-active,  potentially explaining the observed peak.

We have tested this hypothesis in two different ways. The first was to check whether DFT calculations starting with the proposed structure, converge to a structure with a finite $\delta$, or to a $\delta=0$, within the numerical accuracy. We found the latter to be the case, for calculations fully converged in both k-points mesh and cutoff energy, regardless of the value of $U$ in DFT+U, of the pseudopotentials or the DFT flavor used (we checked local, gradient-corrected, and a hybrid functional). 

Next, we calculated the Raman activity of the E$_{2g}$ mode as well as the initially forbidden  B$_{1u}$ mode using the structure suggested in Ref.~\cite{wu2025}, even though it was not the lowest-energy structure in the calculations. We employed the Placzek formalism \cite{Placzek} to compute the Raman  efficiency directly from DFT.

Within the Placzek approximation, the Raman efficiency of a phonon mode is given by \cite{Placzek,cardona2005light}
\begin{equation}
S_{is}(\omega)
\;=\;
V\,(4\pi)^2\left(\frac{\omega}{c}\right)^{4}
\left|
\frac{\partial\varepsilon_{is}(\omega)}{\partial Q}
\right|^{2},
\label{eq:placzek}
\end{equation}
where $V$ is the unit cell volume, $\omega$ is the incident photon frequency, $c$ is the speed of light, and $Q$ is the normal coordinate of the phonon mode. The indices $i$ and $s$ denotes the polarizations of the incident and scattered light.
$\frac{\partial\varepsilon_{is}(\omega)}{\partial Q}$ is the measure of how the phonon displacement modulates the macroscopic dielectric response.

Several features of Eq. (\ref{eq:placzek}) can be summarized as follows.
(1) The $(\omega/c)^4$ prefactor reflects the familiar wavelength dependence of light scattering. 
(2) The Raman activity is entirely encoded in $|\partial\varepsilon/\partial Q|^2$; a mode that does not modulate the dielectric tensor is Raman-silent.
    (3) The polarization indices enforce symmetry selection rules: certain modes appear only for specific incident-scattered polarization combination.
(4) For a mode that becomes active only through a small symmetry-breaking distortion $\delta$, the derivative $\partial \varepsilon/\partial Q$ itself scales as $\delta$, leading to an intensity scaling as $\delta^2$.

\begin{figure}[htb]
    \centering
    \includegraphics[width=0.85\linewidth]{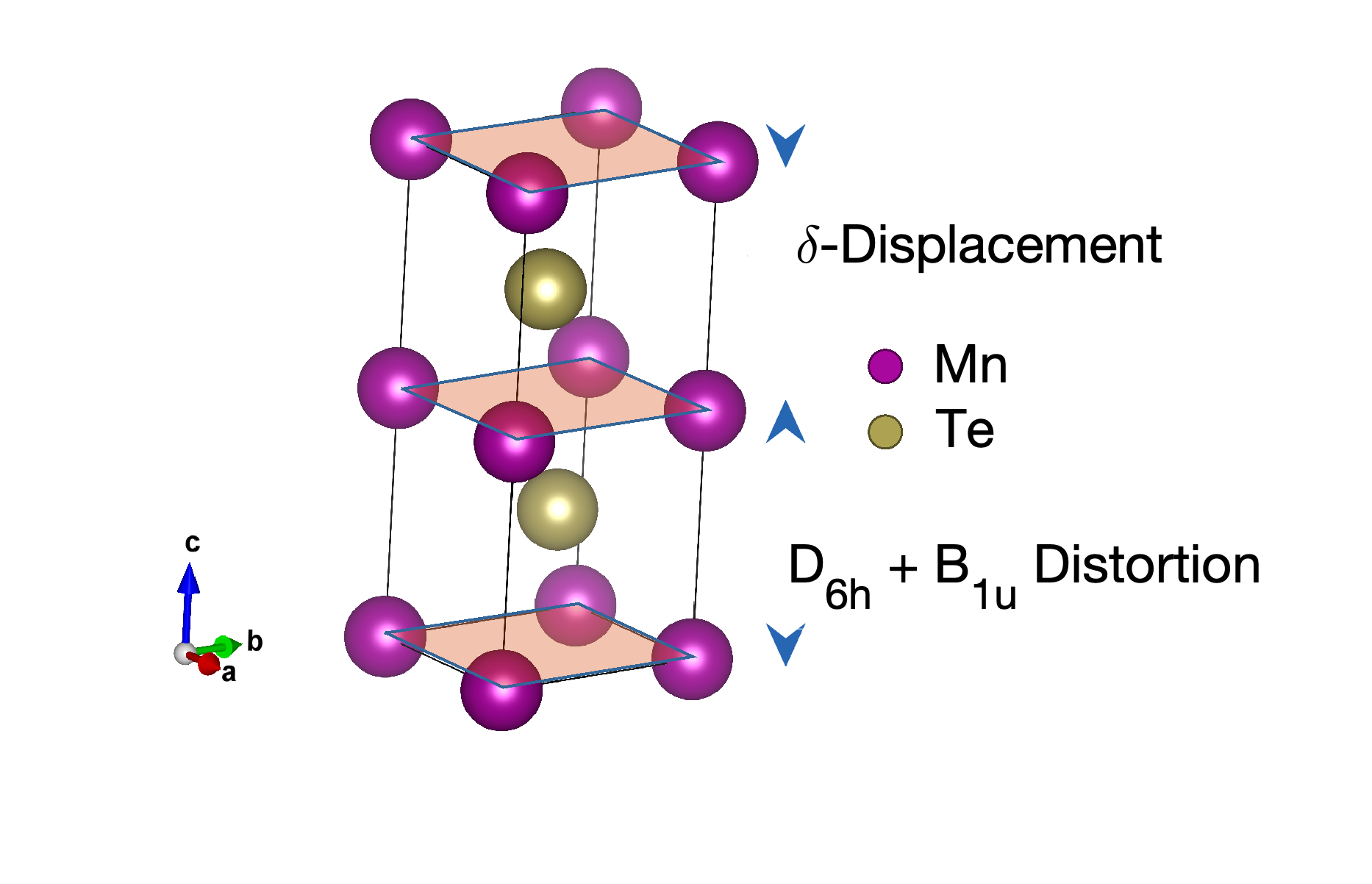}
    \caption{Symmetry lowering $B_{1u}$ mode applied to parent $P6_3/mmc$.}
    \label{fig:D3h}
\end{figure}
By computing $\partial\varepsilon/\partial Q$ for the B$_{1u}$ derived mode in the $P\bar{6}m2$ structure (see Fig.~\ref{fig:D3h}) as a function of $\delta$, we can quantitatively assess whether the proposed symmetry lowering  can produces a measurable Raman intensity.

First-principles calculations were performed within DFT using the Vienna \emph{ab initio} simulation package (VASP) \cite{VASP-1,VASP-2,VASP-3}. 
Generalized gradient approximation
(GGA) \cite{PBE} functional was used.
On-site electronic correlations in Mn $3d$ states were taken into
account using the DFT+$U$ method, with an effective
Hubbard parameter $U_{\mathrm{eff}} = 2.0$~eV. Varying or even eliminating $U$ did not lead to a qualitative change in the results.

A plane-wave cutoff was 540~eV, and  a $\Gamma$-centered $10\times10\times6$ $k$-point mesh was used.
Spin-orbit coupling was not included in the final calculations, as it was found to have a very small effects on the phonon frequencies or dielectric functions in the relevant frequency range.

Starting from the parent $P6_3/mmc$ (No.~194) structure, a $B_{1u}$ symmetry-breaking lattice distortion was frozen in to obtain a reduced-symmetry $P\bar{6}m2$ (No.~187) as shown in Fig.~(\ref{fig:D3h}). 
The phonon eigenvectors $\mathbf e_{n}$ and frequencies $\omega_{\mathrm{ph}}$ (Table~\ref{tab:phonon_modes}) of this  structure were obtained using density-functional perturbation theory. Mode symmetries and selection rules were 
determined using the ISOTROPY software suite \cite{stokessmodes,FINDSYM}.


To test whether the symmetry-breaking distortion proposed in Ref.~\cite{wu2025} produces measurable Raman activity, we constructed  $P\bar{6}m2$ (\#187) structures for various $\delta$.
For each structure the Raman tensor elements were evaluated by displacing atoms along the phonon eigenvector and computing the induced change in the macroscopic dielectric tensor, with atomic displacements  $\mathbf{u}_n=\Lambda\,\mathbf{e}_n$, where $\mathbf{e}_n$ is the normalized phonon eigenvector for atom $n$. 
The displacement amplitude $\Lambda$ was chosen as $\Lambda=A_{\max}/e_{\max}$, with $A_{\max}=0.02$~\AA{}, ensuring numerical accuracy of the finite-difference derivative.

The Raman tensor elements were then obtained by numerical differentiationof the dielectric function $\varepsilon_{\alpha\beta}(\omega_L)$
where $\omega_L$ is the laser frequency ($\hbar\omega_L=1.96$~eV).

For each distorted structure a self-consistent electronic calculation was first performed to obtain the charge density, followed by a non-self-consistent optical calculation with \texttt{LOPTICS = .TRUE.} on a highly refined k-mesh.

The normal-coordinate mode amplitude $Q$ was defined, using atomic units, as usual, as $\mathbf{u}_n=\Lambda\,\mathbf{e}_n$ was taken as
\begin{equation}
Q
=
\Lambda\,\sqrt{2\,\omega_{\mathrm{ph}}}\,
\sum_n \sqrt{M_n}\,\|\mathbf{e}_n\|^2,
\end{equation}
where $\omega_{\mathrm{ph}}$ is the phonon  frequency and $M_n$ are the atomic masses.

\begin{table}[ht]
\centering
\caption{Zone-center optical phonon modes of MnTe in the undistorted 
$P6_3/mmc$ (No.~194) structure and the $P\bar{6}m2$ (No.~187) structure 
obtained by freezing in a $B_{1u}$ symmetry-breaking distortion 
(0.1\% displacement along the $c$ axis). 
 }
\vspace{0.3em}
\begin{tabular}{@{}ccc@{\hskip 1.5em}ccc@{}}
\toprule
\multicolumn{3}{c}{Undistorted ($P6_3/mmc$, \#194)} 
& \multicolumn{3}{c}{Distorted ($P\bar{6}m2$, \#187)} \\
\cmidrule(r){1-3} \cmidrule(l){4-6}
$\omega_{\mathrm{ph}}$ (cm$^{-1}$) & Mode & Activity 
& $\omega_{\mathrm{ph}}$ (cm$^{-1}$) & Mode & Activity \\
\midrule
176.91 & $B_{1u}$  & silent    & 176.43 & $A_1'$      & R \\
126.47 & $E_{1u}$  & IR        & 126.29 & $E'$        & R, IR \\
124.97 & $A_{2u}$  & IR        & 124.76 & $A''_2$     & IR \\
121.74 & $B_{2g}$  & silent    & 121.62 & $A''_2$     & IR \\
 91.27 & $E_{2g}$  & R         & 91.23  & $E'$        & R, IR \\
 90.13 & $E_{2u}$  & silent    & 89.51  & $E''$       & R \\
\bottomrule
\end{tabular}
\vspace{0.5em}

\footnotesize
\textit{Note:} R = Raman active; IR = infrared active. 
The $B_{1u}$ distortion lowers the symmetry and activates 
previously silent modes.
\label{tab:phonon_modes}
\end{table}

\begin{figure}[t]
    \centering
    \includegraphics[width=0.85\linewidth]{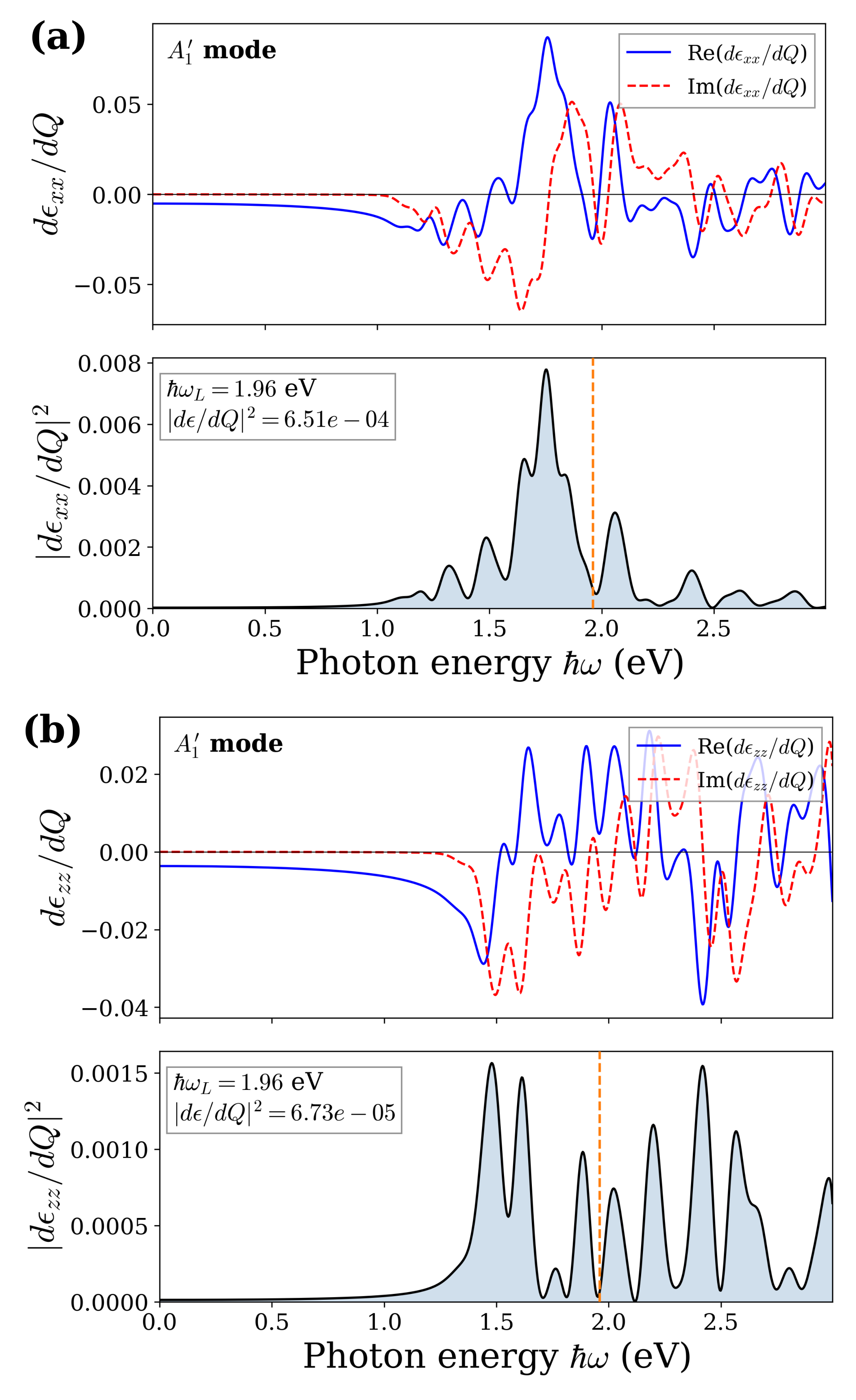}
    \caption{
    Dielectric derivatives $d\varepsilon_{\alpha\beta}/dQ$ for the $A_1'$ Raman mode of MnTe
    ($P\bar{6}m2$, $D_{3h}$).
    \textbf{(a)} In-plane response $d\varepsilon_{xx}/dQ$.
    \textbf{(b)} Out-of-plane response $d\varepsilon_{zz}/dQ$.
    In each panel, the upper subplot shows the real and imaginary parts of
    $d\varepsilon_{\alpha\beta}/dQ$, while the lower subplot shows
    $|d\varepsilon_{\alpha\beta}/dQ|^2$, which is proportional to the Raman intensity. By hexagonal symmetry, $\varepsilon_{xx}=\varepsilon_{yy}$; the cross-polarization components vanish for the totally symmetric $A_1'$ mode, consistent with the experimental observation that this mode is active only in parallel (XX) polarization geometry. The out-of-plane ($zz$) response is
    significantly weaker than the in-plane components.
    }
    \label{fig:phononResponse}
\end{figure}

\begin{table}[ht]
\centering
\caption{Comparison of the calculated Raman response
$|\partial\varepsilon_{\alpha\beta}/\partial Q|^2$
for the $A_1'$ and $E'$ phonon modes in the distorted $P\bar{6}m2$ structure.
The $A_1'$ mode originates from the silent $B_{1u}$ mode and becomes
Raman-active only due to symmetry breaking, whereas the $E'$ mode
(descended from $E_{2g}$) is intrinsically Raman-active.
The dielectric response was evaluated at the incident laser photon energy
$\hbar\omega_L = 1.96$~eV ($\lambda = 633$~nm), corresponding to the
experimental back-scattering Raman configuration.}
\vspace{0.3em}
\begin{tabular}{@{}lcccc@{}}
\toprule
Mode & Origin & $\omega_{\mathrm{ph}}$ (cm$^{-1}$) & $|d\varepsilon_{xx}/dQ|^2$ & $|d\varepsilon_{zz}/dQ|^2$ \\
\midrule
$A_1'$ & $B_{1u}$  & 176 & $\sim 6.51\times 10^{-4}$ & $\sim 6.7\times10^{-5}$ \\
$E'$   & $E_{2g}$  &  91 & $\sim 4.59\times 10^{-2}$   & - \\
\midrule
\multicolumn{3}{r}{Ratio ($E'/A_1'$)} & $\sim 70$ & \\
\bottomrule
\end{tabular}
\vspace{0.3em}
\label{tab:raman_comparison}
\end{table}

The results are summarized in Fig. \ref{fig:phononResponse} and in Table\ref{tab:raman_comparison} (the calculated frequencies are presented in Table\ref{tab:phonon_modes}).
We see that the Raman intensity of the A$_1'$ mode is nearly two orders of magnitude weaker than that of the $E'$ (or $E_{2g}$ in the original structure) and not that much larger than the calculated intensity of the same mode in the $zz$ polarization, where it is forbidden by symmetry. 
Thus, we can conclude, with confidence, 
the distortion-induced activity is way too weak to explain the experimentally observed mode at $\sim\qty{175}{\per\centi\meter}$.

\paragraph{Plasmon hypothesis --}

If this feature is not a phonon, what origin can it have?
There is a Raman-active magnon in this structure\cite{Suszkevicz}, but it has a much lower energy and is only observed well below $T_N$. Electronic charge fluctuations (electronic Raman scattering) \cite{Devereaux-Hackl} remain the only plausible candidate.

Electronic Raman scattering is proportional to $-\mathrm{Im}[1/\epsilon(\omega)]$, and in metals the Raman frequency is well below the plasma frequency where this function peaks. Still, even in good metals it is quite observable and routinely used to determine the superconducting gap, for instance. It is hard to compare the absolute intensities of electronic and phononic Raman scattering in  DFT calculations, but it is worth mentioning that, to a good approximation (specifically, resonant effects), the prefactor for the Raman intensity is determined by the fluctuations of the inverse effective mass over the Fermi surface, i.e., $m^{-1}(\mathbf{k})-\langle m^{-1}\rangle$, where  $m^{-1}(\mathbf{k})$ is the second derivative of electron energy with respect to momentum, and $\langle m^{-1}\rangle$ is its average  over the Fermi surface \cite{Devereaux-Hackl}. As discussed below, at the relevant doping level several bands with very different effective masses contribute to transport, assuring that the prefactor is not small.

Let us now estimate the frequency of such plasmon. Pristine MnTe is a $p$-type semiconductor with a typical hole concentration on the order $p\sim10^{18}$ cm$^{-3}$, or $10^{-4}$ per unit cell (Table \ref{table3}). The bands near the valence band maximum at the A point \cite{Osumi2024,Faria2023} are dominated by the Te $p_x$ and $p_y$ orbital character and well described, in the altermagnetically ordered state, by a 4-band $\mathbf{k}\cdot\mathbf{p}$ Hamiltonian including symmetry-allowed spin-orbit coupling (SOC) up to linear order in $k$ \cite{Belashchenko2025}. We will use this $\mathbf{k}\cdot\mathbf{p}$ model to describe the transport properties of MnTe. Without SOC, the bands in this model are fourfold degenerate along the $A\Gamma$ direction with the out-of-plane effective mass $m^*_\perp=1.14m_0$, while the in-plane dispersion exhibits two pairs of degenerate bands: light holes with $m^*_{lh}=0.13m_0$ and heavy holes with $m^*_{hh}=0.53m_0$. Spin-orbit coupling lifts the degeneracy at a generic $k$ and induces conical dispersions near the A point. The A point itself remains fourfold degenerate up to a tiny splitting in the order of 1 meV, which may be ignored for our purposes.

\begin{figure}[htb]
    \centering
    \includegraphics[width=0.85\linewidth]{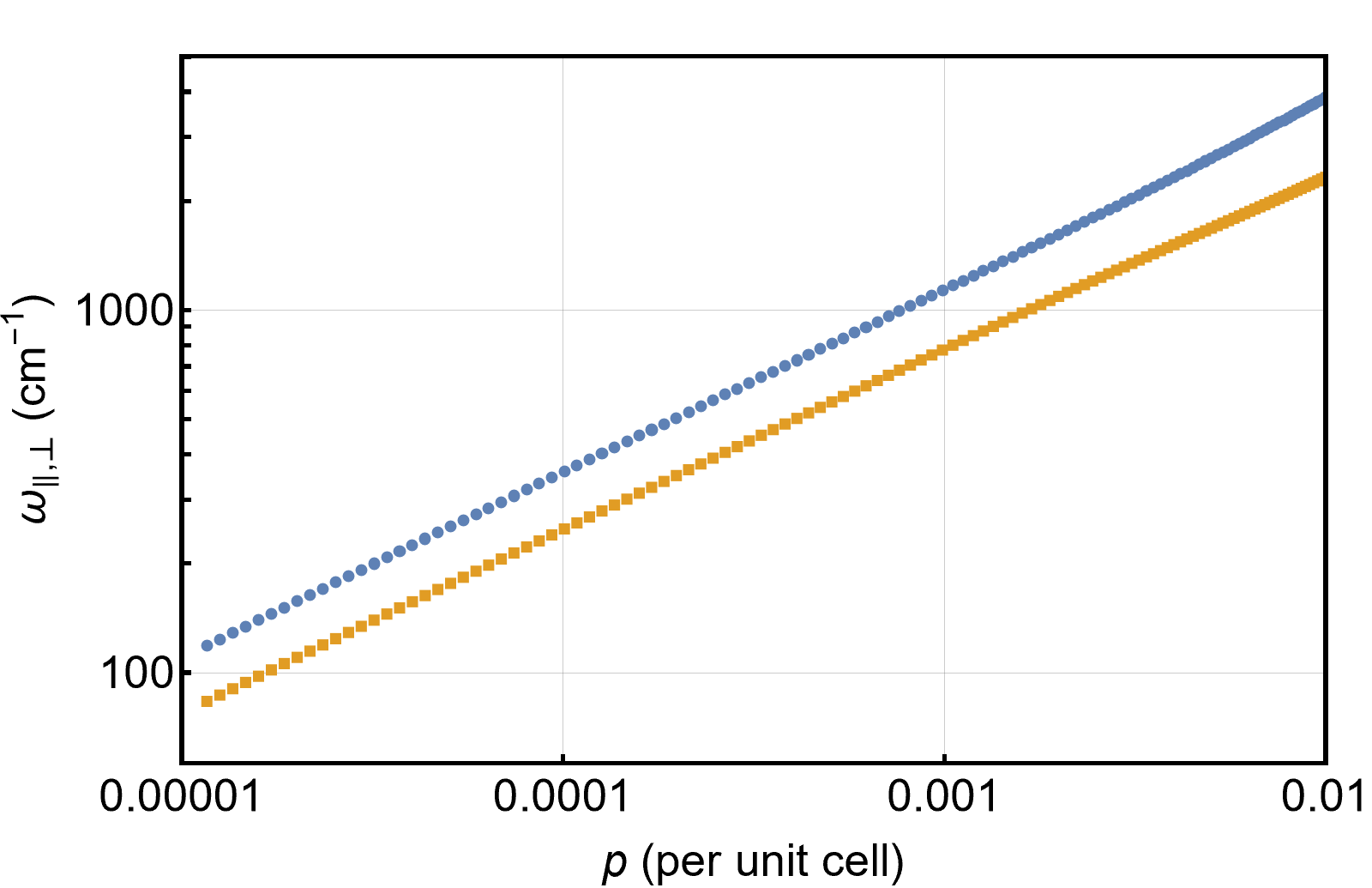}
    \caption{Plasma frequencies $\omega_\parallel$ (blue circles) and $\omega_\perp$ (orange squares) as a function of the hole concentration $p$ at $T=150$ K, calculated assuming $\varepsilon=\varepsilon_0$.}
    \label{fig:plasmafreq}
\end{figure}

The tensor of squared plasma frequency can be calculated directly from the band structure using the standard definition:
\begin{align}
    (\omega^2_p)_{\alpha\beta}=\varepsilon^{-1}\sum_n\int v_{n\alpha} v_{n\beta}\frac{\partial f(E_n)}{\partial \mu}\frac{d^3k}{(2\pi)^3},
    \label{omegap2}
\end{align}
where $\varepsilon$ is the dielectric constant in the absence of charge carriers, the $\mathbf{v}_n(\mathbf{k})$ and $E_n(\mathbf{k})$ are the group velocity and energy of band $n$, and $f(E_n)$ is the Fermi-Dirac distribution at chemical potential $\mu$. The two independent components of this tensor in hexagonal MnTe are $\omega^2_{p,xx}=\omega^2_\parallel$ and $\omega^2_{p,zz}=\omega^2_\perp$. The resulting $\omega_\parallel$ and $\omega_\perp$ are plotted in Fig. \ref{fig:plasmafreq} as a function of the hole concentration $p$, {neglecting semiconducting screening} for the moment.  {Reported values in Table \ref{table3} cluster around 2--7$\times 10^{18}$ cm$^{-1}$, corresponding in Fig. \ref{fig:plasmafreq} to  
$\omega_\parallel\sim$ 500--1000 cm$^{-1}$ and $\omega_\perp\sim 350$--660 cm$^{-1}$.
Now, these should be divided by $\sqrt{\varepsilon(0)}$, the refraction index of the undoped MnTe. Our calculations give $\varepsilon_c(0)\approx\varepsilon_{ab}(0)\approx 10$, in agreement with previous calculations and experiments (see Ref. \cite{Sano2025}, Fig. 4). Thus, we expect the plasmon frequencies to be between 170--320 and 120--220 cm$^{-1}$, respectively, in the ballpark of the observed Raman line. Given a semiquantative character of the estimates above, the plasmon origin is plausible.
}

The above calculation assumes that transport in MnTe is well described by the band model, which requires that the mean-free path $\lambda$ is at least a few times larger than the interatomic distance. We estimate $\lambda$ using the experimental data on the electric resistivity \cite{Wasscher-thesis} and the $\mathbf{k}\cdot\mathbf{p}$ band structure.

In the relaxation-time approximation, the in-plane conductivity is $\sigma_\parallel=\varepsilon\omega^2_{\parallel}\tau$. According to Ref. \onlinecite{Wasscher-thesis}, the resistivity of undoped MnTe samples at $T=150$ K (which is well below the N\'eel temperature $T_N$) varies approximately from 0.03 to 3 \unit{\ohm\centi\meter}, depending on the heat treatment and crystallographic orientation of the sample. Using our calculated $\sqrt{\varepsilon}\omega_{\parallel}\approx 360$ cm$^{-1}$, the corresponding range of the relaxation time is $\tau\sim0.8$--80 \unit{\femto\second}.

The mean squared group velocity was found by dividing $\varepsilon\omega^2_\parallel$ from Eq. (\ref{omegap2}) by the similar density-of-states-like integral without the velocities in the integrand. This results in $\sqrt{\langle v_x^2\rangle}\approx\qty{3.7e5}{\meter/\second}$. The mean-free path can then be estimated as $\lambda=\sqrt{2\langle v_x^2\rangle}\,{\tau}\sim\numrange[range-phrase={\text{--}}]{0.3}{30}~\unit{\nano\meter}$. 

Thus, with the possible exception of the most resistive samples with $\rho\gtrsim\qty{1}{\ohm\centi\meter}$, the mean-free path corresponding to the experimental resistivity exceeds the lattice spacing, supporting the treatment of transport properties within the band model. This conclusion is consistent with the temperature dependence $\rho(T)$ of the resistivity in those samples \cite{Wasscher-thesis}: those with $\rho<\qty{1}{\ohm\centi\meter}$
have metallic (monotonically increasing) $\rho(T)$, while those with $\rho<\qty{1}{\ohm\centi\meter}$ have a minimum at $T\sim100$ K, suggesting thermally-activated carriers with activation energy on the order of \qty{10}{\milli\electronvolt}.


\paragraph{Conclusions.}
From our analysis, we arrive at the following conclusions:\\
1. 
The strong Raman-active mode observed at about \qty{175}{\per\centi\meter} and routinely assigned to the $E_{2g}$ phonon is not the latter (as suggested by Wu \emph{et al.} \cite{wu2025}), and is a Raman excitation of unknown origin. \\
2. 
This excitation, contrary to the conjecture of Ref. \cite{wu2025}, is not the forbidden $B_{1u}$ mode that ``leaks'' because of an intrinsic symmetry lowering.\\
3. In principle, one can never exclude an impurity phase that would happen to have a strong A$_g$ mode in the right energy range. However, the \qty{175}{\per\centi\meter} mode is rather strong in most experiments, and none of those papers reported a sizable impurity phase. This is a question that should be answered by future experiments with well-controlled compositions. \\
4. A plausible explanation is that this mode is a plasmon excitation due to hole self-doping. {There are open questions with this interpretation as well --- for instance, why the second plasmon is not visible: too soft? too weak? too broad? However, it is clear from our results that this mode is not a phonon, and the frequencies of the hole-doped plasmons are in the right ballpark. Rephrasing Holmes, ``When you have eliminated all which is impossible, then whatever remains, however improbable, is worth considering''.}

Final verification of this hypothesis will be provided by future experiments, and can shed new light on the nature of doping and electronic transport (including, but not limited to altermagnetic transport effects) in MnTe.

\paragraph*{Experimental outlook ---} Possible experimental tests of the plasmon hypothesis include direct detection of plasmons via energy loss spectroscopy or electronic microscopy, accessing the out-of-plane polarized plasmon, which should appear at a different  energy, using the corresponding light polarization, as well as indirect tests using MnTe modified in a way that does not significantly change the phonon properties but does affect the carrier concentration, such as very low-dosage doping with Li. 

 \begin{acknowledgments}We acknowledge helpful discussion with Thomas Devereaux, Rudi Hackl, Patrick Vora, Alberto de la Torre, Badih Assaf, Angela Hight-Walker and Rafa\"el Hermann.
The work at George Mason University was supported by the Army Research Office under Cooperative Agreement Number W911NF-22-2-0173, and that at the University of Nebraska by the U.S. Department of Energy (DOE) Established Program to Stimulate Competitive Research (EPSCoR) through Grant No. DE-SC0024284. Most calculations were performed  using resources provided by the Office of Research Computing at George Mason University (URL: https://orc.gmu.edu), funded in part by grants from the National Science Foundation (Award Number 2018631).
\end{acknowledgments}
\bibliography{bib}
\end{document}